\documentstyle[epsf,onecolumn]{jpsj_sissa}
\newcommand{\pr}{Phys.\ Rev. }
\newcommand{\prb}{\pr B }
\newcommand{\prl}{\pr Lett. }
\newcommand{\rmp}{Rev.\ Mod.\ Phys. }
\newcommand{\jpsj}{J.\ Phys.\ Soc.\ Jpn. }

\newcommand{\Tr}{{\rm Tr}}
\newcommand{\Tc}{T_{\rm c}}
\newcommand{\rmc}{{\rm c}}
\newcommand{\rme}{{\rm e}}
\newcommand{\romap}{{\rm p}}
\newcommand{\rms}{{\rm s}}

\newcommand{\veps}{\varepsilon}
\newcommand{\bfveps}{\mib{\veps}}
\newcommand{\vepsz}{\veps^{(0)}}
\newcommand{\vepskin}{\veps_{\rm kin}}

\newcommand{\vepscor}[1]{\veps_{{\rm cor}#1}}
\newcommand{\Ekin}{E_{\rm kin}}

\newcommand{\bfchi}{\mib{\chi}}
\newcommand{\DeltaMH}{\Delta^{\rm MH}}
\newcommand{\muz}{\mu^{(0)}}
\newcommand{\bfSigma}{\mib{\Sigma}}

\newcommand{\ho}{{\hat{\omega}}}
\newcommand{\delho}{\delta\ho}
\newcommand{\ave}[1]{{\langle#1\rangle}}
\newcommand{\com}[2]{\left[#2\right]_{#1}\!}
\newcommand{\up}{\uparrow}
\newcommand{\down}{\downarrow}

\newcommand{\Psid}{\Psi^\dagger}

\newcommand{\bfPsi}{{\bf\Psi}}
\newcommand{\bfPsid}{\bfPsi^\dagger}

\title{Operator Projection Method Applied
to the Single-Particle Green's Function in the Hubbard Model}
\def\runtitle{Operator Projection Method Applied
to the Single-Particle Green's Function in the Hubbard Model}

\author{Shigeki Onoda and Masatoshi Imada}
\def\runauthor{Shigeki Onoda and Masatoshi Imada}

\inst{Institute for Solid State Physics, University of Tokyo,
Kashiwanoha 5-1-5, Kashiwa, Chiba 277-8581, Japan}
\recdate{}

\abst{A new non-perturbative framework for many-body correlated systems
is formulated by extending the operator projection method (OPM).
This method offers a systematic expansion which enables us to project
into the low-energy structure after extracting the higher-energy hierarchy.
This method also opens a way to systematically take into account the effects
of collective excitations.
The Mott-Hubbard metal-insulator transition in the Hubbard model
is studied by means of this projection beyond the second order
by taking into account magnetic and charge fluctuations
in the presence of the high-energy Mott-Hubbard structure.
At half filling, the Mott-Hubbard gap is correctly reproduced
between the separated two bands.
Near half filling, strongly renormalized low-energy single-particle
excitations coexisting with the Mott-Hubbard bands are shown to appear.
The significance of the momentum-dependent self-energy in the results is stressed.
}

\kword{metal-insulator transition, Hubbard model, projection operator}

\begin{document}
\sloppy
\maketitle

In spite of a long history~\cite{Mott,Hubbard,Kanamori63,RMP_Imada},
a satisfactory theoretical description
of strongly correlated electrons remains open because of the difficulty
in reproducing their rich hierarchy structures in energy and momentum.
The properties near the metal-insulator transitions (MITs) provide
a typical challenge.
It is required to correctly describe both a high-energy incoherent structure
of the "Hubbard bands" with the Mott gap
and the low-energy coherent excitations.
The low-energy part must be severely renormalized
with strong momentum dependence.
It is also required to treat magnetic and charge fluctuations.

The dynamical mean field theory (DMFT)~\cite{MetznerVollhardt89,RMP_DMFT}
reproduces both the incoherent Hubbard bands and the quasiparticle structure.
On the other hand, numerical simulations~\cite{RMP_Imada} have revealed
an important aspect of the MIT in the two-dimensional Hubbard model,
i.e., nearly dispersionless fermions around the $(\pi,0)$ and $(0,\pi)$
momenta~\cite{HankePreuss,Assaad}.
This suggests that the momentum dependence of the single-particle self-energy,
which is completely neglected in the DMFT, is significant
in real finite-dimensional systems.
The momentum dependence of the low-energy structure appears
from collective modes under the influence of high-energy incoherence.
Such hierarchy can not be properly treated in the perturbative or one-loop
analyses of the self-energy and the dynamical susceptibilities,
because the high-energy structure such as the Mott-Hubbard bands is missing.

In this letter, we propose a systematic expansion which incorporates
both the coherent and incoherent structure in the energy hierarchy
of single-particle excitations.
By applying the operator projection method (OPM) to the single-particle Green's function,
our framework allows us to systematically project out the high-energy structure
and sequentially extract the low-energy coherent excitations.
The low-energy coherence appears with the strongly momentum dependent
self-energy due to two-particle collective fluctuations.

The OPM first developed by Nakajima~\cite{Nakajima58}, Zwanzig~\cite{Zwanzig}
and Mori~\cite{Mori65} is based on the equation of motion~\cite{Zubarev60}
with a moment expansion of a correlation function.
The main procedure of this method is to decompose the time-derivative
of an operator into the term projected onto itself
and the remaining new operator.
The OPM involves repeatedly applying this procedure
to the newly created operators.
Then, the OPM allows one to construct a continued-fraction expansion
of a correlation function. The expansion is suited for systematically
evaluating the energy hierarchy structure from high to low energies.
Similar equation-of-motion (EQM) approaches supplemented
with various decoupling schemes and/or other approximations
have been intensively studied for strongly correlated models
such as the Hubbard~\cite{Hubbard,Roth69,BeenenEdwards52,MatsumotoMancini97}
and the $t$-$J$ models~\cite{Plakida,MatsumotoSaikawaMancini96,Prelovsek}.
For the Hubbard model, the Hubbard bands are reproduced
from the first three, i.e., the zeroth-, first- and second-order moments
of the Green's function~\cite{Hubbard}.
In principle, the low-energy band appears in the fourth order,
although a systematic and correct expansion has not yet been carried out.
In this letter, we show that the OPM up to the second order,
supplemented with the self-energy in the second-order projection
which is referred to as the second-order self-energy,
makes it possible to correctly reproduce the hierarchy:
the Hubbard bands as well as the renormalized low-energy bands.
We note that our calculated second-order self-energy
gives an evaluation of the fourth-order moment.
The OPM based on the equation of motion is a non-perturbative treatment
and works from the weakly correlated limit
$U\ave{n}(2-\ave{n})/t\to0$ to the strongly correlated limit
$U\ave{n}(2-\ave{n})/t\to\infty$.

We take the Hubbard Hamiltonian with an electron transfer
$t_{\vec{x},\vec{x}'}$ from an atomic site $\vec{x}'$ to $\vec{x}$
and the local Coulomb repulsion $U$;
$H=-\sum_{\vec{x},\vec{x}',s}t_{\vec{x},\vec{x}'}
c^\dagger_{\vec{x}s}c_{\vec{x}'s}
+U\sum_{\vec{x}}n_{\vec{x}\up}n_{\vec{x}\down}$
with $n_{\vec{x}s}=c^\dagger_{\vec{x}s}c_{\vec{x}s}$, or
$\bar{H}=H-\mu\sum_{\vec{x}s}n_{\vec{x}s}$ with the chemical potential $\mu$.
We perform the projection procedure for the electron creation and annihilation
operators at an atomic site $\vec{x}$ with a spin index $s$,
$c^\dagger_{\vec{x},s}$ and $c_{\vec{x},s}$.
It is useful to define the fermionic $4N$-component vector operator composed of
\begin{equation}
\Psi_{\vec{x}}=
\left(\sigma_0\frac{\rho_0+\rho_3}{2}-i\sigma_2\frac{\rho_0-\rho_3}{2}\right)
{}^t(c_{\vec{x}\up},c_{\vec{x}\down},c^\dagger_{\vec{x}\up},
c^\dagger_{\vec{x}\down}).
\label{eq:Psi_x}
\end{equation}
Here, $N$ is the total number of the atomic sites.
$\bfPsid$ is defined as the Hermitian conjugate operator of $\bfPsi$.
$\sigma_0$ ($\rho_0$) and $\vec{\sigma}$ ($\vec{\rho}$) are the identity
and the Pauli matrices, respectively,
which operate to the spin (charge) space of vector operators as
$\Psi$ from the left-hand side and those as $\Psid$ from the right-hand side.

We introduce $\ho A=[A,\bar{H}]_-$,
and $\ave{A}=\Tr[e^{-\bar{H}/T}A]/\Tr[e^{-\bar{H}/T}]$
at temperature $T$.
We also define the response function of $\bfPsi$
in the $4N\times4N$ matrix representation $\mib{K}_{\Psi,\Psid}(t)$.
Its $(a,a')\otimes(\vec{x},\vec{x}')$ component is defined by
\begin{equation}
K_{\Psi,\Psid}^{aa'}(t;\vec{x},\vec{x}')
=-i\ave{\com{+}{\Psi_{\vec{x}}^a(t),(\Psid){}_{\vec{x}'}^{a'}}}.
\label{eq:K_Psi,Psid}
\end{equation}
We note $\mib{K}_{\Psi,\Psid}(0)=-i\mib{I}$
with the $4N\times4N$ identity matrix $\mib{I}$
because of the fermion anticommutation relation.
The Fourier transformation of Eq. (\ref{eq:K_Psi,Psid})
gives the susceptibility $\bfchi_{\Psi,\Psid}(\omega)$,
which is nothing but the Green's function $\mib{G}(\omega)$.
The projection procedure is defined as
\begin{subequations}
\begin{eqnarray}
\ho\bfPsi&=&\bfveps^{(11)}\bfPsi+\delho\bfPsi,
\slabel{eq:1EQM-G:ho-Psi}\\
\bfveps^{(11)}&=&\mib{K}_{\ho\Psi,\Psid}(0)\mib{K}^{-1}_{\Psi,\Psid}(0).
\slabel{eq:1EQM-G:veps11}
\end{eqnarray}
\label{eq:1EQM-G:project1-Psi}
\end{subequations}
Namely, the operator $\ho\bfPsi$ is split into $\bfPsi$ with a component
given by the equal-time correlation $\mib{K}_{\ho\Psi,\Psid}(0)$
and the new operator $\delho\bfPsi$ which satisfies
$\mib{K}_{\Psi,(\delho\Psi)^\dagger}(0)=0$ by definition.
It allows us to express the single-particle Green's function
$\mib{G}(\omega)=\bfchi_{\Psi,\Psid}(\omega)$ in the form of a Dyson equation
as
\begin{eqnarray}
\mib{G}(\omega)
&=&i\left[\mib{G}^{(0)}{}^{-1}(\omega)-\bfSigma_1(\omega)\right]^{-1}
\mib{K}_{\Psi,\Psid}(0),
\label{eq:1EQM-G:Dyson-G}\\
\mib{G}^{(0)}(\omega)&=&\left[\omega\mib{I}-\bfveps^{(11)}\right]^{-1},
\label{eq:1EQM-G:chi_Psi^0}\\
\bfSigma_1(\omega)
&=&-i\bfchi_{\delho\Psi,(\delho\Psi)^\dagger}^{\rm irr}(\omega)
\mib{K}_{\Psi,\Psid}^{-1}(0).
\label{eq:1EQM-G:Dyson-chi_Psi}
\end{eqnarray}
The irreducible part of $\bfchi_{\delho\Psi,(\delho\Psi)^\dagger}(\omega)$
with respect to $\mib{G}^{(0)}(\omega)$ has been introduced;
\begin{eqnarray}
\lefteqn{\bfchi_{\delho\Psi,(\delho\Psi)^\dagger}^{\rm irr}(\omega)}
\nonumber\\
&=&\left[\bfchi_{\delho\Psi,(\delho\Psi)^\dagger}{}^{-1}(\omega)
-i\mib{K}_{\Psi,\Psid}^{-1}(0)\mib{G}^{(0)}(\omega)\right]^{-1}.
\label{eq:1EQM-G:chi_delho-Psi_delho-Psid}
\end{eqnarray}
For the Hubbard model, one obtains
\begin{equation}
\veps^{(11)}_{\vec{x},\vec{x}'}
=\vepsz_{\vec{x},\vec{x}'}\sigma_0\rho_3-U\delta_{\vec{x},\vec{x}'}
\left[\ave{\vec{S}_{\vec{x}}}\cdot\vec{\sigma}\rho_0
-\frac{\ave{\Delta^{\rms(s)\bar{i}}_{\vec{x}}}}{\sqrt{2}}
\sigma_0\rho_{\bar{i}}\right],
\label{eq:1EQM-G:veps11_x}
\end{equation}
with $\vepsz_{\vec{x},\vec{x}'}=-t_{\vec{x},\vec{x}'}
-(\mu-\frac{U}{2}\ave{n_{\vec{x}}})\delta_{\vec{x},\vec{x}'}$,
$n_{\vec{x}}=n_{\vec{x}\up}+n_{\vec{x}\down}$,
$S^j_{\vec{x}}=c^\dagger_{\vec{x}s}\sigma^j_{ss'}c_{\vec{x}s'}/2$,
$\Delta^{\rms(s)1}_{\vec{x}}=\frac{1}{\sqrt{2}}(c_{\vec{x}\up}c_{\vec{x}\down}
+c^\dagger_{\vec{x}\down}c^\dagger_{\vec{x}\up})$ and
$\Delta^{\rms(s)2}_{\vec{x}}=\frac{i}{\sqrt{2}}(c_{\vec{x}\up}c_{\vec{x}\down}
-c^\dagger_{\vec{x}\down}c^\dagger_{\vec{x}\up})$
The summation over $\bar{i}=1$ and $2$ should be taken.
For the repulsively interacting model ($U>0$),
we can exclude the superconducting states with the isotropic $s$-wave
pairing symmetry, $\ave{\Delta^{\rms(s)i}_{\vec{x}}}=0$.
The thermal average $\ave{\vec{S}_{\vec{x}}}$, if it does not vanish,
is chosen to break the symmetry in the $z$-axis,
i.e., $\ave{S^{\pm}_{\vec{x}}}=0$.
Then, Eq. (\ref{eq:1EQM-G:veps11_x}) and $\delho\bfPsi$ are reduced to
\begin{subequations}
\begin{eqnarray}
\veps^{(11)}_{\vec{x},\vec{x}'}
&=&\vepsz_{\vec{x},\vec{x}'}\sigma_0\rho_3
-U\ave{S^z_{\vec{x}}}\delta_{\vec{x},\vec{x}'}\sigma_3\rho_0,
\label{eq:veps11_x_2}\\
\delho\Psi_{\vec{x}}&=&U\left(\frac{1}{2}\delta n_{\vec{x}}\sigma_0\rho_3
-\delta S^z_{\vec{x}}\sigma_3\rho_0\right)\Psi_{\vec{x}},
\label{eq:1EQM-G:delhho-Psi_x}
\end{eqnarray}
\label{eq:1EQM-G:project1-Psi_x_2}
\end{subequations}
with $\delta n_{\vec{x}}=n_{\vec{x}}-\ave{n_{\vec{x}}}$ and
$\delta S^z_{\vec{x}}=S^z_{\vec{x}}-\ave{S^z_{\vec{x}}}$.

If $\bfSigma_1(\omega)$ is ignored, then the present theory is reduced
to a Hartree-Fock theory. In order to calculate $\bfSigma_1(\omega)$,
one can adopt perturbation expansions on a single-particle basis,
one-loop approximation, long-time or short-time approximations.
However, if $\bfSigma_1(\omega)$ is self-consistently determined
by means of simple frameworks such as perturbation expansion or
one-loop approximations~\cite{Baym62,KadanoffBaym,FLEX,Bickers91},
the MIT at half filling can not be reached~\cite{TPSC}.
This is because arguments along this line always fail to give
the correct high-energy behavior of the self-energy~\cite{TPSC}.
We note that only the zeroth- and first-order moments of the Green's function
are correct in these theories.
As known in the Hubbard and two-pole approximations, we need to proceed
to higher-order moments, as is discussed below.

The second-order projection gives
\begin{subequations}
\begin{eqnarray}
\ho\delho\bfPsi
&=&\bfveps^{(21)}\bfPsi+\bfveps^{(22)}\delho\bfPsi+\delho\delho\bfPsi,
\label{eq:2EQM-G:ho-delho-Psi}\\
\bfveps^{(21)}&=&\mib{K}_{\ho\delho\Psi,\Psid}(0)\mib{K}^{-1}_{\Psi,\Psid}(0),
\label{eq:2EQM-G:veps21}\\
\bfveps^{(22)}&=&\mib{K}_{\ho\delho\Psi,(\delho\Psi)^\dagger}(0)
\mib{K}^{-1}_{\delho\Psi,(\delho\Psi)^\dagger}(0).
\label{eq:2EQM-G:veps22}
\end{eqnarray}
\label{eq:2EQM-G:project2-Psi}
\end{subequations}
From Eq. (\ref{eq:2EQM-G:project2-Psi}) one obtains the irreducible self-energy
\begin{subequations}
\begin{eqnarray}
\hspace*{-20pt}\bfSigma_1(\omega)&=&\left[\bfSigma_1^{(0)}{}^{-1}(\omega)
-\bfSigma_2(\omega)\right]^{-1}\bfveps^{(21)},
\label{eq:2EQM-G:Sigma_1}\\
\hspace*{-20pt}\bfSigma_1^{(0)}(\omega)&=&\left[\omega\mib{I}-\bfveps^{(22)}\right]^{-1},
\label{eq:2EQM-G:Sigma_1^0}\\
\hspace*{-20pt}\bfSigma_2(\omega)
&=&-i\bfchi^{\rm irr}_{\delho\delho\Psi,(\delho\delho\Psi)^\dagger}(\omega)
\mib{K}^{-1}_{\delho\Psi,(\delho\Psi)^\dagger}(0).
\label{eq:2EQM-G:Sigma_2}
\end{eqnarray}
\label{eq:2EQM-G:Sigma_1,Sigma_2}
\end{subequations}
For the repulsive Hubbard model,
\begin{subequations}
\begin{eqnarray}
\veps^{(21)}_{\vec{x},\vec{x}'}&=&U^2\delta_{\vec{x},\vec{x}'}
\ave{\left(\delta n_{\vec{x}}\sigma_0\rho_3/2
-\delta S^z_{\vec{x}}\sigma_3\rho_0\right)^2},
\label{eq:2EQM-G:veps21_x}\\
\veps^{(22)}_{\vec{x},\vec{x}'}&=&-t^{(22)}_{\vec{x},\vec{x}'}\rho_3
-\left(\mu_2\sigma_0\rho_3-\ave{S^z_x}\sigma_3\rho_0\right)
\delta_{\vec{x},\vec{x}'},
\label{eq:2EQM-G:veps22_x}\\
t^{(22)}_{\vec{x},\vec{x}'}&=&t_{\vec{x},\vec{x}'}
\langle\left(\delta n_{\vec{x}}\sigma_0\rho_3/2
-\delta\vec{S}_{\vec{x}}\cdot\vec{\sigma}\rho_0
+\Delta^{\rms(s)\bar{i}}_{\vec{x}}\sigma_0\rho_{\bar{i}}/\sqrt{2}\right)
\nonumber\\
&&{}\times\left(\delta n_{\vec{x}'}\sigma_0\rho_3/2
-\delta\vec{S}_{\vec{x}'}\cdot\vec{\sigma}\rho_0
-\Delta^{\rms(s)\bar{i}'}_{\vec{x}'}\sigma_0\rho_{\bar{i}'}/\sqrt{2}\right)\rangle
\nonumber\\
&&{}\times\ave{\left(\delta n_{\vec{x}}\sigma_0\rho_3/2
-\delta S^z_{\vec{x}}\sigma_3\rho_0\right)^2}{}^{-1},
\label{eq:2EQM-G:t^2_x}\\
\mu_2&=&\muz_2\sigma_0\rho_0+\frac{(1-\ave{n})\vepskin+\vepscor{2}}
{\ave{(\delta n_{\vec{x}}\sigma_0\rho_3/2
-\delta S^z_{\vec{x}}\sigma_3\rho_0){}^2}},
\label{eq:2EQM-G:mu2}\\
\vepskin&=&-\frac{1}{N}\sum_{\vec{k}}\!t_{\vec{k}}\!
\left(\ave{n_{0;\vec{k}}}\rho_3\sigma_0/2
-\ave{S^z_{0;\vec{k}}}\rho_0\sigma_3\right),
\label{eq:2EQM-G:vepskin}
\end{eqnarray}
\label{eq:2EQM-G:project2-Psi_x}
\end{subequations}
\noindent
where $\muz_2=\mu-U(1-\frac{1}{2}\ave{n_{\vec{x}}})$.
The kinetic energy per site summed over spins is expressed as
$\Ekin=\frac{1}{2}\Tr\vepskin(\rho_0+\rho_3)=
-\frac{1}{N}\sum_{\vec{k},s}t_{\vec{k}}\ave{c^\dagger_{\vec{k}s}c_{\vec{k}s}}$.
Below, a momentum independent energy shift $\vepscor{2}$
due to two-site correlated hopping terms which take the form
$t_{\vec{x},\vec{\bar{x}}}\ave{c^\dagger_{\vec{x}s}c_{\vec{\bar{x}}s'}
\delta n_{\vec{x}-s}}$ in Eq. (\ref{eq:2EQM-G:veps22_x}) are neglected.
Then, in the case of $\ave{S^z_{\vec{x}}}=0$, the remaining operator becomes
\begin{eqnarray}
\lefteqn{\hspace*{-20pt}\delho\delho\Psi_{\vec{x}}
\approx-U\left(\frac{1}{2}\delta n_{\vec{x}}\rho_3\sigma_0
-\delta\vec{S}_{\vec{x}}\cdot\vec{\sigma}\rho_0
+\frac{\Delta^{\rms(s)\bar{i}}_{\vec{x}}}{\sqrt{2}}\rho_{\bar{i}}\sigma_0\right)}
\nonumber\\
&&\hspace*{-20pt}\rho_3t_{\vec{x},\vec{\bar{x}}}\Psi_{\bar{\vec{x}}}
+\left[\frac{4(1-\ave{n})}{\ave{n}(2-\ave{n})}
\delta_{\vec{x},\vec{\bar{x}}}\vepskin
+t^{(22)}_{\vec{x},\vec{\bar{x}}}\right]\rho_3\delho\Psi_{\vec{\bar{x}}}.
\label{eq:2EQM-G:delho-delho-Psi_x}
\end{eqnarray}

Even when $\bfSigma_2(\omega)$ is neglected, the present formalism is
a generalization beyond the Hubbard I approximation~\cite{Hubbard},
because not only the Mott-Hubbard bands are reproduced but also
$\bfveps^{(22)}$ depends on the momentum~\cite{Roth69,BeenenEdwards52}.
The former fact is guaranteed,
since the Pauli exclusion principle is satisfied.
The latter fact modifies the single-particle dispersion
in each Mott-Hubbard band.
The momentum dependence of $\bfveps^{(22)}$ enters the present theory
through the equal-time charge, spin, isotropic $s$-wave pairing correlations,
while these correlations are ignored in the Hubbard I approximation~\cite{Hubbard}.
In the symmetry-unbroken phase,
the electron Green's function with momentum $\vec{k}$ and spin $s$ reads
\begin{equation}
G_{\rme,s}(\omega,\vec{k})=\left[\omega-\veps^{(0)}_{\vec{k}}
-\frac{\left(\frac{U}{2}\right)^2\ave{n}(2-\ave{n})}
{\omega-\tilde{\veps}_{\vec{k}}}\right]^{-1},
\label{eq:2EQM-G:G_2^0}
\end{equation}
with $\tilde{\veps}_{\vec{k}}=-\tilde{t}_{\vec{k}}-\muz_2
-\frac{2(1-\ave{n})\Ekin}{\ave{n}(2-\ave{n})}$.
Here, $\tilde{t}_{\vec{k}}$ is a (1,1) component of Eq. (\ref{eq:2EQM-G:t^2_x}).
Equation (\ref{eq:2EQM-G:G_2^0}) has a pole at $\omega^\pm_{\vec{k}}
=(\vepsz_{\vec{k}}+\tilde{\veps}_{\vec{k}}\pm\DeltaMH_{\vec{k}})/2$
in each Hubbard band with a residue $z^\pm_{\vec{k}}
=(1\pm(\vepsz_{\vec{k}}-\tilde{\veps}_{\vec{k}})/\DeltaMH_{\vec{k}})/2$.
Here, $\DeltaMH_{\vec{k}}=\sqrt{U^2\ave{n}(2-\ave{n})
+(\vepsz_{\vec{k}}-\tilde{\veps}_{\vec{k}})^2}$ is the momentum-dependent
Mott-Hubbard gap.
The momentum dependence diminishes as $U/t$ diverges.
It is easily found that when the collective degrees of freedom are completely
frozen, i.e., $\tilde{t}_{\vec{k}}=0$, Eq. (\ref{eq:2EQM-G:G_2^0})
recovers the Hubbard I solution at half filling.
On the other hand,
when equal-time two-particle correlations with the momentum $\vec{Q}$
are significant, $\tilde{t}_{\vec{k}}$ behaves like $t_{\vec{k}+\vec{Q}}$.
Then, Eq. (\ref{eq:2EQM-G:G_2^0}) smoothly connects
the symmetry-broken and symmetry-unbroken phases through the growth of
short-range correlations in Eq. (\ref{eq:2EQM-G:t^2_x}).
For example, it reproduces the shadow bands in systems with well-developed
short-range antiferromagnetic correlations.
It can also describe other kinds of shadow bands
such as nearly ferromagnetic systems.
So far, few theories have reproduced such shadow-band effect
even qualitatively within the one-loop approximations to
$\bfSigma_1(\omega)$~\cite{TPSC,TPSC2}.
Without $\bfSigma_2(\omega)$, the expectation value of the Hamiltonian
as a function of $\tilde{t}_{\vec{k}}$ is given by
\begin{eqnarray*}
\lefteqn{\ave{H}=T\!\sum_{\omega_n,\vec{k},s}\!
\left[-t_{\vec{k}}+\frac{U\ave{n}}{2}+\Sigma_{1\rme,s}(\omega_n,\vec{k})\right]
G_{\rme,s}(\omega_n,\vec{k})}
\\
&=&2\sum_{\vec{k},\pm}\!\left[
z^\pm_{\vec{k}}\left(-t_{\vec{k}}+\frac{U\ave{n}}{2}\right)
\pm\frac{U^2\ave{n}(2-\ave{n})}{4\DeltaMH_{\vec{k}}}\right]
f(\omega^\pm_{\vec{k}}),
\end{eqnarray*}
with a fermionic Matsubara frequency $\omega_n=(2n+1)\pi T$
and the Fermi distribution function $f(\omega)=(e^{\omega/T}+1)^{-1}$.
The momentum-$\vec{k}$ spin-$s$ component of $\bfSigma_1(\omega)$
has been introduced as $\Sigma_{1\rme,s}(\omega,\vec{k})$.
At $\ave{n}=1$, we obtain $\ave{H}=-\sum_{\vec{k}}[t_{\vec{k}}^2
-(t_{\vec{k}}+\tilde{t}_{\vec{k}})^2/4]/U$ when $U\gg T$, $t$.
This includes the superexchange term through Eq. (\ref{eq:2EQM-G:t^2_x}).

Next, we consider the second-order self-energy $\bfSigma_2(\omega)$
by introducing decoupling approximations.
Here, we can take several methods to calculate $\bfSigma_2(\omega)$:
(A) the perturbation theory, (B) one-loop approximation
with two-particle susceptibilities obtained by the self-consistent RPA
(SCRPA)~\cite{SCRPA},
and (C) one-loop approximation with those obtained
from the two-particle self-consistent (TPSC) method~\cite{TPSC,TPSC2}.
In the TPSC method, irreducible vertices are determined so that
the local-moment sum rule is also fulfilled for the two-particle properties,
not only charge and spin correlations~\cite{TPSC}
but also local $s$-wave pairing correlations~\cite{TPSC2}.
Therefore, we take the last method, (C).
We note that $\bfSigma_2(\omega)$ has been evaluated within a two-site level
by means of a different decoupling scheme~\cite{MatsumotoMancini97}.
However, short-ranged correlations are ignored there.

Then, the second-order electronic self-energy
is evaluated by means of the decoupling approximation as
\begin{eqnarray}
\lefteqn{\Sigma_{2\rme}(\omega_n,\vec{k})=\frac{2T/N}{\ave{n}(2\!-\!\ave{n})}\!
\sum_{m,\vec{q}}\left[G(\omega_n-\Omega_m,\vec{k}-\vec{q})
\right.}\nonumber\\
&&\left\{\frac{1}{2}T_{\vec{k};\vec{q}}^2
\left(\chi_\rmc(\Omega_m,\vec{q})+\chi_\rms(\Omega_m,\vec{q})\right)
+t_{\vec{k}-\vec{q}}^2\chi_\rms(\Omega_m,\vec{q})\!\right\}\!
\nonumber\\
&&\left.-2t_{\vec{k}-\vec{q}}^2
G(\Omega_m-\omega_n,\vec{q}-\vec{k})\chi_\romap(\Omega_m,\vec{q})\right],
\label{eq:2EQM-G:Sigma_2_decoupling}
\end{eqnarray}
where $T_{\vec{k};\vec{q}}=t_{\vec{k}-\vec{q}}
-\frac{2(1-\ave{n})}{\ave{n}(2-\ave{n})}\Ekin-t^{(22)}_{\vec{k}}$,
$\Omega_m=2\pi mT$ is a bosonic Matsubara frequency,
and $\chi_\rmc$, $\chi_\rms$ and $\chi_\romap$ are the charge, spin,
local $s$-wave pairing susceptibilities, respectively.
The spin indices have not been included since all the variables are
independent of the spin index.
Within the TPSC approximation, the susceptibilities are calculated from
\begin{subequations}
\begin{eqnarray}
\chi_\rmc(\Omega_m,\vec{k})\!&=&\!2\chi_{\rm ph}(\Omega_m,\vec{k})
/[1+\Gamma_\rmc\chi_{\rm ph}(\Omega_m,\vec{k})],
\label{TPSC:chin}\\
\chi_\rms(\Omega_m,\vec{k})\!&=&\!2\chi_{\rm ph}(\Omega_m,\vec{k})
/[1-\Gamma_s\chi_{\rm ph}(\Omega_m,\vec{k})],
\label{TPSC:chis}\\
\chi_\romap(\Omega_m,\vec{k})\!&=&\!\chi_{\rm pp}(\Omega_m,\vec{k})
/[1+\Gamma_\romap\chi_{\rm pp}(\Omega_m,\vec{k})],
\label{TPSC:chip}
\end{eqnarray}
\end{subequations}
with an approximation
$\Gamma_\rms=4U\ave{n_{\vec{x}\up}n_{\vec{x}\down}}/\ave{n}^2$
and the self-consistency conditions for $\Gamma_\rms$, $\Gamma_\rmc$
and $\Gamma_\romap$ given by the local moment sum rules;
\begin{subequations}
\begin{eqnarray}
&&\hspace*{-20pt}T/N\sum{}_{m,\vec{k}}\chi_\rmc(\Omega_m,\vec{k})
=\ave{n}+2\ave{n_{\vec{x}\up}n_{\vec{x}\down}}-\ave{n}^2,
\\
&&\hspace*{-20pt}T/N\sum{}_{m,\vec{k}}\chi_\rms(\Omega_m,\vec{k})
=\ave{n}-2\ave{n_{\vec{x}\up}n_{\vec{x}\down}},
\\
&&\hspace*{-20pt}T/N\sum{}_{m,\vec{k}}\chi_\romap(\Omega_m,\vec{k})
=\ave{n_{\vec{x}\up}n_{\vec{x}\down}}.
\end{eqnarray}
\end{subequations}
$\chi_{\rm ph}$ and $\chi_{\rm pp}$ are the particle-hole and particle-particle susceptibilities calculated from the bare Green's functions.

\begin{figure}[tbh]
\begin{center}
\epsfxsize=6.5cm
$$\epsffile{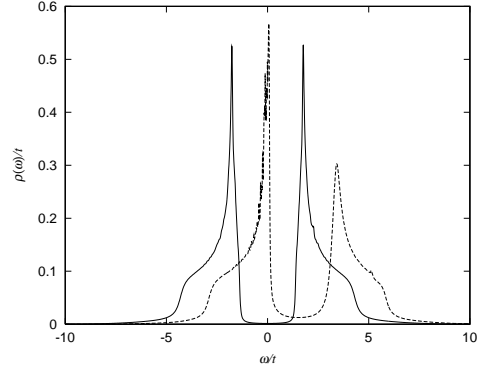}$$
\vspace*{-12mm}
\caption{The local density of states for $U=4t$ and $T=0.02t$
at $n=1$ (solid line) and $n=0.9$ (dashed line).}
\label{fig:DOS}
\end{center}
\end{figure}
\begin{figure}[tbh]
\begin{center}
\vspace*{-17pt}
\epsfxsize=7.0cm
$$\epsffile{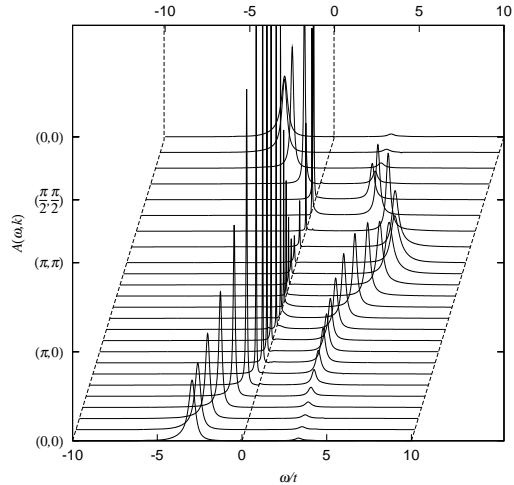}$$
\vspace*{-12mm}
\caption{Single-particle spectral weights $A(\omega,\vec{k})=-{\rm Im}G_{\rme,s}(\omega,\vec{k})/\pi$ for $U=4t$ and $T=0.02t$ at $n=0.9$.}
\label{fig:Akw}
\end{center}
\end{figure}
The calculated density of states for the Hubbard model
with $U=4t$ and the electron transfer restricted to the nearest neighbor
at the fillings $n=0.9$ and $n=1$ are shown in Fig.~\ref{fig:DOS}.
The calculations have been done in the $32\times32$ lattices
with $2048$ Matsubara frequencies.
The metal and the Mott insulator are reproduced at $n=0.9$ and $n=1$,
respectively, with the Hubbard bands.
The emergence of a flat dispersion around $(\pi,0)$ is in agreement
with the numerical results~\cite{Assaad}.
It is remarkable that at $n=0.9$, low-energy peaks
around the magnetic Brillouin zone lie below the chemical potential
and the dispersion outside the zone becomes strongly small
and pinned near $\omega=0$, as shown in Fig.~\ref{fig:Akw},
which agrees with previous results~\cite{HankePreuss,MatsumotoMancini97}.
It turns out that the growth of the equal-time antiferromagnetic
(AFM) spin correlation produces
such strong modification of the dispersion in the scale of $\tilde{t}$.
At $n=1$, $\tilde{t}/t$ should become $-1$ at the zero temperature
to yield the antiferromagnetism with the broken translational symmetry.
In the present calculation carried out at $T=0.02t$, in fact,
$\tilde{t}/t=-0.84$ for $n=1$, while $\tilde{t}/t=-0.53$ for $n=0.9$.
This indicates that as holes are doped into the Mott insulator,
the shadow-band dispersion moves toward the chemical potential.
However, self-consistent calculations of $\bfSigma_2(\omega)$
using the TPSC method tend to weaken the SDW shadow-band structure too much
and to overestimate the low-energy flat bands, as shown in Fig.~\ref{fig:Akw},
because the TPSC method overestimates $\Sigma_{2\rme}(\omega,\vec{k})$
at low frequencies.
This will be improved by considering higher-order projections,
as is discussed elsewhere.

We also note that in the results calculated from the (A) or (B) method,
the AFM spin correlations appear to be underestimated.
These give qualitatively similar results but smaller values of $|\tilde{t}|$.

In summary, the OPM has been applied to the two-dimensional Hubbard model.
Beyond the second-order projection, which correctly reproduces the first four
moments and yields the two-pole approximation~\cite{Roth69},
the higher-order dynamics in the self-energy have been
self-consistently taken into account.
Then, we obtained not only the Mott insulator at half filling
but also nearly dispersionless low-energy excitations pinned
near the Fermi level away from half filling.
This is the first analytic theory which has succeeded in describing
both of the Mott-Hubbard bands and the low-energy single-particle excitations
near half filling under strong short-ranged spin correlations.
The OPM applied to the two-particle susceptibilities
beyond the SCRPA~\cite{SCRPA} are now under study
to give more correct two-particle properties.
Higher-order projections are necessary to consider
the non-local pairing correlations, which are crucial
in relation to the high-$\Tc$ superconductivity~\cite{Onoda}.

The authors thank P. Prelovsek for illuminating discussions at the initial stage
of this work.
The work was supported by the "Research for the Future" program
of the Japan Society for the Promotion of Science under grant number
JSPS-RFTF97P01103.

\end{document}